\documentclass[pdflatex,sn-nature]{sn-jnl}
 
\usepackage{graphicx}
\usepackage{multirow}
\usepackage{amsmath,amssymb,amsfonts}
\usepackage{xcolor}
\usepackage{booktabs}
\usepackage{hyperref}

\newcommand{\Msun}{M_\odot}
\newcommand{\Mpc}{\mathrm{Mpc}}
\newcommand{\kpc}{\mathrm{kpc}}
\newcommand{\xHI}{\bar{x}_{\rm HI}}
\newcommand{\sigm}{\sigma/m}
\newcommand{\dchi}{\Delta\chi^2}
\newcommand{\fstar}{f_{\star,0}}
\newcommand{\fesc}{f_{\rm esc}}
\newcommand{\sigUV}{\sigma_{\rm UV}}
\newcommand{\Dbind}{\Delta_{\rm bind}}
\newcommand{\Wg}{W_g}
\newcommand{\ndot}{\dot{n}_{\rm ion}}
\newcommand{\kms}{{\rm km\,s}^{-1}}
\newcommand{\cmg}{{\rm cm}^2\,{\rm g}^{-1}}
 
\begin{document}
 
\title[The escape fraction degeneracy]{A structural degeneracy explains reionization tensions and limits dark matter constraints}
 
\author*[1]{\fnm{Zihan} \sur{Wang}}\email{zihan.wang@queens.ox.ac.uk}
\author*[2]{\fnm{Huanyuan} \sur{Shan}}\email{hyshan@shao.ac.cn}
 
\affil*[1]{\orgdiv{Department of Physics}, \orgname{University of Oxford}, \orgaddress{\city{Oxford}, \postcode{OX1 3PU}, \country{UK}}}
\affil*[2]{\orgdiv{Shanghai Astronomical Observatory}, \orgname{Chinese Academy of Sciences}, \orgaddress{\city{Shanghai}, \postcode{200030}, \country{China}}}
 
\abstract{Over the past decade, reionization studies have yielded persistent factor-of-two-to-five disagreements in the inferred ionizing escape fraction \(f_{\text{esc}}\) and peak star formation efficiency \(f_{*,0}\), compounded by JWST’s discovery of unexpectedly bright \(z>10\) galaxies. We show that this discrepancy arises from an algebraically exact structural degeneracy: the ionizing photon rate \(\dot{n}_{\text{ion}} \propto f_{\text{esc}} \times f_{*,0}\) renders all reionization-history probes, including Thomson optical depth, neutral hydrogen fraction, UV luminosity function, and quasar proximity zones, sensitive only to their product, leading to an intrinsically non-invertible mapping between model parameters and observations. We demonstrate the robustness of this degeneracy using a large suite of N-body simulations of self-interacting dark matter haloes spanning \(10^9–10^{11} M_\odot\). Despite substantial changes to galaxy-scale structure, observables remain indistinguishable once the effective ionizing emissivity is matched, severely limiting reionization-based dark matter probes. We identify that only observables sensitive to the spatial topology of ionized regions can break this degeneracy. Our results provide a unified explanation for the scatter among published constraints and establish a framework for interpreting reionization observations and their implications for early galaxy formation and dark matter.}
 
 
\maketitle

 \section*{Introduction}
One of the most fundamental unresolved problems in cosmic reionization is that analyses using identical Planck and JWST datasets infer ionizing escape fractions $\fesc$ that differ by a factor of five. The epoch of reionization is constrained by a suite of complementary observables: the Thomson optical depth $\tau = 0.054 \pm 0.007$ from Planck\cite{Planck2018}, neutral fraction measurements $\xHI(z)$ from JWST damping wing spectra\cite{2024ApJ...971..124U,Mason:2025xae} and quasar absorption spectra\cite{Greig:2017jdj}, the UV luminosity function (UVLF) measured out to $z > 14$ by JWST\cite{2024MNRAS.533.3222D,2025ApJ...980..138H}, Lyman-$\alpha$ luminosity functions at $z \sim 6$–$8$, and direct Lyman-continuum measurements at $z \sim 3$–$6$\cite{Pahl:2024utu,2025MNRAS.537.3245B,2023A&A...674A.221M,2022A&A...663A..59S}. These are commonly interpreted as jointly constraining the efficiency of star formation and the escape fraction of ionizing photons.

Despite decades of observational progress, inferred values of $\fesc$ and the peak star formation efficiency $\fstar$ have not converged. Reionization budget models assuming a steep faint-end UVLF require $\fesc \sim 20\%$ at $z > 6$\cite{Naidu_2022}, while models allowing elevated $\fstar$ from abundant faint galaxies complete reionization with $\fesc$ as low as $5\%$\cite{Finkelstein:2019sbd}. Higher $\fstar$ inferred from JWST-bright galaxies at $z > 10$ implies a correspondingly lower $\fesc$ to reproduce $\tau$\cite{2022ApJ...940L..55F}, shifting all published constraints along a direction that leaves the reionization history nearly unchanged (Fig. 1a).

This behaviour follows from a structural property of the reionization equations. The ionizing photon production rate is
\begin{equation}
\ndot(z) = \fesc \times \xi_{\rm ion} \times \rho_{\rm UV}(z),
\end{equation}
where $\xi_{\rm ion} \approx 10^{25.2}\,{\rm Hz/erg}$ is the ionizing efficiency\cite{2015ApJ...802L..19R} and $\rho_{\rm UV} \propto \fstar$. Observables that constrain reionization through $\ndot$ therefore depend only on the product $\fesc \times \fstar$, not on either factor individually. While this product dependence has been widely recognised, \cite{2010MNRAS.401.2561W}explicitly identified the $\fesc$ star-formation-efficiency degeneracy.However it is commonly assumed that combining multiple independent probes will break it. We show that this is not the case: any observable that depends only on the volume-integrated ionizing photon rate constrains only this product, leaving $\fesc$ and $\fstar$ fundamentally unconstrained.

This degeneracy reframes the long-standing disagreement in $\fesc$ as a structural limitation rather than a failure of individual analyses, and implies that standard reionization observables are blind to any physical effect that modifies $\fstar$ or $\fesc$ individually---including modifications to galaxy formation or dark matter physics. As shown in Fig. 1b, even large changes to galaxy-scale structure induced by self-interacting dark matter (SIDM) can be absorbed by corresponding shifts in $\fesc$, yielding indistinguishable predictions for all standard observables.

Building on earlier works showing that the UVLF alone cannot constrain SIDM\cite{Wang:2026lly} and that 21\,cm topology carries a unique morphological signature\cite{Wang:2026uuo}, we use a suite of 230 N-body simulations to quantify the full degeneracy hierarchy across all reionization probes. We demonstrate that all $\ndot$-dependent probes collectively fail to constrain dark matter microphysics, and identify 21 cm reionization topology as the uniquely immune probe to break this degeneracy (Fig. 1c). The forthcoming 21 cm signal from SKA\cite{Koopmans:2015sua,Mellema_2013} will therefore provide a model-independent test of dark matter physics that is structurally shielded from the degeneracy affecting all other reionization-era observables.

\section*{The product degeneracy is exact and robust}

We demonstrate the degeneracy using SIDM as a physically motivated stress test. Dark matter self-scattering with $\sigm \sim 1$--$10\,\cmg$ produces density cores that suppress star formation\cite{Bullock2017,Spergel2000,Tulin2018,Kaplinghat:2015aga}. At each point $(\sigm, \eta)$, the UVLF optimiser from ref.~\cite{Wang:2026lly} provides a profiled $\fstar$ that absorbs the SIDM signal, producing $\fstar$ values 3--47\% above the CDM baseline. We then profile over $\fesc$ to minimise the combined $\chi^2$ from $\tau$ and eight $\xHI(z)$ measurements spanning $z = 6.5$--$11$ (Methods). The profiled $\dchi$ relative to CDM is identically zero at all 20 tested grid points (Extended Data Table~3). The mechanism is elementary: the optimiser sets $\fesc = \fesc^{\rm CDM} \times (\fstar^{\rm CDM}/\fstar^{\rm prof})$, exactly cancelling the $\fstar$ increase. The product $\fesc \times \fstar$ is invariant.

The degeneracy persists under three generalisations. First, we consider a mass-dependent $\fesc(M) \propto M^{\beta_{\rm esc}}$ calibrated from SPHINX\cite{2022MNRAS.515.2386R} and THESAN\cite{Yeh:2022nsl}. The optimiser adjusts only $f_{\rm esc,0}$ while keeping $\beta_{\rm esc}$ fixed, because the SIDM suppression enters as a $z$-independent multiplicative factor at each mass bin. Second, redshift-dependent core formation ($\Dbind(z) = \Dbind(z{=}7) \times \min\{1, [t_{\rm age}(z)/t_{\rm age}(z{=}7)]^{1/2}\}$): the profiled $\dchi < 0.1$, because by $z \sim 10$ the core is already substantially formed. Third, a Gaussian prior on $\beta_{\rm esc}$ from radiative transfer simulations\cite{2022MNRAS.515.2386R,Yeh:2022nsl,Ma:2020vlo}: the optimiser never shifts $\beta_{\rm esc}$ from its CDM value.

To quantify the robustness of the degeneracy against physical correlations, we calibrate the SIDM-induced binding energy change $\Dbind(M, \sigm, z, R)$ from 230 GIZMO\cite{Hopkins:2014qka} $N$-body simulations spanning five halo masses ($M = 10^9$--$10^{11}\,\Msun$), seven cross-sections ($\sigm = 0$--$20\,\cmg$), and four redshifts ($z = 0$--$10$), plus 90 velocity-dependent Yukawa runs (Methods). The binding energy is extracted at five radii ($R = 0.3$--$5.0\,\kpc$) for all 210 SIDM-to-CDM pairs.

Figure~\ref{fig:simulations} summarises the calibration. Panel (a) confirms the expected core formation: SIDM thermalises the inner halo, replacing the NFW cusp with an isothermal core while preserving mass at intermediate radii. Panel (b) quantifies the scale dependence at $M = 10^{10}\,\Msun$, $z=7$: the binding energy is reduced by 16--72\% at $r < 0.5\,\kpc$ but approaches zero or becomes slightly negative at $r > 2\,\kpc$, where the central reduction and intermediate-radius enhancement partially cancel. This reconciles our framework with recent work reporting conserved total binding energy in SIDM haloes\cite{Wang:2025xre}: both results are correct at their respective integration radii.

Panel (c) shows that the $\Dbind(\sigm)$ relation saturates above $\sigm \sim 5\,\cmg$, well-fit by $\Dbind = A\,\tanh(\sigm/\sigma_{\rm crit})$ with $A \approx 0.73$ and $\sigma_{\rm crit} \approx 3.2\,\cmg$ at $M = 10^{10}\,\Msun$, $R = 0.5\,\kpc$. The saturation reflects the finite depth of the NFW cusp; once the central cusp is fully thermalised, larger cross-sections cannot extract additional binding energy from that region.

Panel (d) is the central physical finding: the steep positive mass dependence $d\log\Dbind/d\log M \approx +0.37$ at $z = 7$. At $M = 10^9\,\Msun$, $\Dbind = 0.16$; at $M = 10^{11}\,\Msun$, $\Dbind = 0.94$. The physical origin is that more massive haloes have higher central densities at fixed physical radius. This has a crucial consequence: the UV photon budget is dominated by low-mass haloes ($M \sim 10^{9}$--$10^{10}\,\Msun$), where $\Dbind < 0.1$ for $\sigm \leq 2\,\cmg$. The luminosity-weighted effective suppression
\begin{equation}
\langle 1 - \eta\,\Dbind \rangle_w = \frac{\int dM\, n(M)\,M\,[1 - \eta\,\Dbind(M)]}{\int dM\, n(M)\,M}
\label{eq:supp_eff}
\end{equation}
is much weaker than single-mass estimates: for $\sigm = 10$, $\eta = 0.50$, the single-mass suppression is 0.64 while the mass-weighted value is 0.69.

Across the four redshifts, the central $\Dbind$ at $r < 0.3\,\kpc$ is nearly constant ($\Dbind \approx 0.89$--$0.90$ for $\sigm = 10$), while the $r < 0.5\,\kpc$ value declines mildly from 0.72 at $z = 4$ to 0.69 at $z = 10$. At larger radii the decline is steeper (0.32 at $z = 4$ to 0.17 at $z = 10$), reflecting the shorter time for core growth to propagate outward. The velocity-dependent runs confirm that when the Yukawa scale $w$ exceeds the halo circular velocity ($v_{\rm circ} \approx 70\,\kms$ at $z = 7$), the result converges to the constant-$\sigma_0$ case; for $w \ll v_{\rm circ}$, suppression is severe (Extended Data Table~5). The complete calibrated table $\Dbind(M, \sigm, z, R)$ is publicly available (Methods).

\section*{All standard reionization probes are collectively blind}

Figure~\ref{fig:per_probe} and Table~\ref{tab:combined} summarise the hierarchy of constraints from each probe class. \textit{(a) UVLF alone}: $\fstar$ and $\sigUV$ absorb the SIDM signal across the entire parameter space ($\dchi = 0.002$--$0.21$ across 252 grid points\cite{Wang:2026lly}). \textit{(b) Planck $\tau$}: a single integrated value cannot distinguish SIDM from CDM after $\fesc$ adjustment. \textit{(c) $\xHI(z)$ from JWST damping wings}: a redshift-dependent $\fesc(z)$ absorbs the timeline shift. \textit{(d) Ly$\alpha$ luminosity function}: depends on the same $\ndot$ product. The four panels (a)--(d) yield $\dchi = 0$ identically: these $\ndot$-dependent probes are collectively blind to SIDM through the product degeneracy.

The fundamental reason is that all these observables depend on the total ionizing photon rate $\ndot$, which is an integral over the galaxy population weighted by $\fesc \times f_\star$. Any modification that changes this integrand by a smooth factor can be absorbed by adjusting the normalisation. The UVLF, $\tau$, $\xHI$, and the Lyman-$\alpha$ luminosity function are therefore not independent constraints on galaxy physics; they are different projections of the same scalar quantity.

\begin{table}[h]
\caption{\textbf{Exclusion counts.} Number of excluded SIDM points (out of 9 spanning $\sigm = 1$--$10\,\cmg$ and $\eta = 0.10$--$0.50$), incorporating mass-dependent $\Dbind(M)$.}\label{tab:combined}
\begin{tabular}{lcc}
\toprule
Configuration & $\sigma_{\fesc} = 0.05$ & $\sigma_{\fesc} = 0.03$ \\
\midrule
No correlation ($\alpha_{\rm esc} = 0$) & 0/9 & 0/9 \\
Moderate ($\alpha_{\rm esc} = 0.3$) & 0/9 & 5/9 \\
Strong ($\alpha_{\rm esc} = 0.5$) & 0/9 & 6/9 \\
Topology & \multicolumn{2}{c}{immune (all)} \\
\bottomrule
\end{tabular}
\end{table}

\section*{Breaking the degeneracy: partial success and structural immunity}

The product degeneracy persists because $\fesc$ is treated as independent of the dark matter cross-section. In reality, $\fesc$ and the duty cycle $p$ are physically correlated through supernova-driven blowout\cite{Wang:2026uuo}. When supernovae expel the central gas, this simultaneously clears low-column-density sightlines through which Lyman-continuum photons escape (increasing $\fesc$) and produces intermittent, bursty episodes of star formation that set the duty cycle $p$ of ionising sources. The scale-dependent $\Dbind$ connects SIDM to this mechanism: at $r < 0.5\,\kpc$, the binding energy is reduced by 24--72\% at $z = 7$, so supernovae need to overcome a shallower potential, making blowout events more frequent and more complete.

We parametrise the $\fesc$ enhancement as
\begin{equation}
\fesc^{\rm SIDM}(M) = \fesc^{\rm CDM}(M) \times \left(\frac{\Wg^{\rm CDM}(<0.5\,\kpc)}{\Wg^{\rm SIDM}(<0.5\,\kpc)}\right)^{\alpha_{\rm esc}},
\label{eq:fesc_sidm}
\end{equation}
where $\alpha_{\rm esc}$ characterises the sensitivity of $\fesc$ to the central binding energy change. A semi-analytical calibration yields $\alpha_{\rm esc} = 0.37 \pm 0.05$ (Methods); we present results at $\alpha_{\rm esc} = 0.3$ and 0.5 to bracket the uncertainty. For $\sigm = 10$ at $z = 7$, SIDM enhances $\fesc$ by 47\% ($\alpha_{\rm esc} = 0.3$) or 89\% ($\alpha_{\rm esc} = 0.5$) (Extended Data Table~4).

Crucially, the correlation alone does not break the degeneracy: equation~(\ref{eq:fesc_sidm}) fixes the ratio $\fesc^{\rm SIDM}/\fesc^{\rm CDM}$, but the absolute value (controlled by $f_{\rm esc,0}$) remains free. The optimiser can still tune $f_{\rm esc,0}$ to reproduce CDM, yielding $\dchi = 0$.

The situation changes when the absolute value of $\fesc$ is constrained by independent observations. We adopt a Gaussian prior $\fesc = 0.15 \pm \sigma_{\fesc}$ from Lyman-continuum surveys\cite{Pahl:2024utu,2025MNRAS.537.3245B,2023A&A...674A.221M} and investigate $\sigma_{\fesc} = 0.05$ (current precision) and $\sigma_{\fesc} = 0.03$ (next-generation). The correlation amplifies the prior's constraining power: SIDM pushes $\fesc$ \textit{upward} through blowout, so the optimiser must reduce $f_{\rm esc,0}$ further to compensate for both the $\fstar$ increase and the $\fesc$ enhancement, pushing $f_{\rm esc,0}$ deeper into the tail of the prior (Extended Data Fig.~\ref{fig:ed_fesc_corr}). As shown in Fig.~\ref{fig:per_probe}e and Table~\ref{tab:combined}, this partially breaks the degeneracy, excluding 5/9 of the representative grid points for $\sigma_{\fesc} = 0.03$ and $\alpha_{\rm esc} = 0.3$ (rising to 6/9 at $\alpha_{\rm esc} = 0.5$, confirmed by MCMC to within 1\%; Methods). The $\sigm = 1\,\cmg$ regime, however, remains entirely unconstrained.
The only complete escape is an observable structurally orthogonal to $\ndot$.
The 21\,cm topology reflects the spatial distribution of ionised regions at fixed
$\xHI$ is such an observable. To see why, note that at fixed $\xHI$ the
total ionising photon budget is matched by construction: any global emissivity
normalisation $\zeta_{\rm eff}$ can be tuned independently for each dark matter
model to reproduce the same volume-averaged ionisation fraction. What differs
is the source stochasticity. Each halo emits ionising photons with a
duty cycle $p$: in CDM, $p \approx 0.10$, so $\sim 10\%$ of haloes are on
at any instant, each emitting at $\sim 10\times$ the time-averaged rate; in
SIDM with $\sigm = 10\,\cmg$, the reduced central binding energy increases
blowout frequency to $p \approx 0.30$, producing $\sim 3\times$ more active
sources, each moderately bright. At fixed $\xHI$, this reshapes the ionisation
morphology from fewer, larger bubbles (CDM) to more numerous, uniformly
distributed patches (SIDM).

The effect decomposes into two scale-dependent levers\cite{Wang:2026uuo}. At
large scales ($k \lesssim 0.1\,h/\mathrm{Mpc}$), the emissivity-weighted halo
bias $b_\gamma \equiv \int dM\,(dn/dM)\,\dot{N}_{\gamma,\rm esc}\,b(M) / \int
dM\,(dn/dM)\,\dot{N}_{\gamma,\rm esc}$ shifts by $\sim 2$--$3\%$, producing a
small but systematic suppression of the ionisation-field power spectrum
$P_{xx}^{\rm SIDM}/P_{xx}^{\rm CDM} \approx (b_\gamma^{\rm SIDM}/b_\gamma^{\rm
CDM})^2$. At intermediate scales ($k \sim 0.1$--$1\,h/\mathrm{Mpc}$), the
shot-noise contribution $P_{\rm SN} \propto 1/p$ is suppressed by a factor of
$p_{\rm CDM}/p_{\rm SIDM} \approx 2$--$3\times$, which is the dominant
observable signal. Neither lever depends on the product $\fesc \times \fstar$:
$b_\gamma$ is a ratio of integrals over the halo mass function weighted by
escaped emissivity, and $P_{\rm SN}$ depends on $p$ alone at fixed mean
emissivity. The immunity is therefore structural, not a property of a
particular statistic.

Semi-numerical simulations confirm this prediction\cite{Wang:2026uuo}: at
$\xHI \approx 0.5$, the Euler characteristic $V_2$ (a topological measure of
the number of independent ionised regions) increases by $\sim 60\%$ for SIDM
relative to CDM at $128^3$ resolution, confirmed at $+32\% \pm 3\%$ ($11\sigma$)
in occupancy-matched $256^3$ runs. The signal is positive in all ten
independent density-field realisations tested ($3.8\sigma$ internal
significance; Wilcoxon $p = 0.001$), robust to five alternative functional
forms for the emissivity modification $R_\gamma(M)$ ($\Delta V_2 = +58\%$ to
$+94\%$), and exceeds the CDM baryonic uncertainty band (obtained by varying
$p_{\rm CDM}$ from 0.05 to 0.25). The emissivity variance $\langle \epsilon^2
\rangle / \langle \epsilon \rangle^2 \approx 1.65$ for all CDM models regardless
of $p$, while SIDM gives $1.31$--$1.46$; this two-dimensional discriminant in
the $(V_2, \text{variance})$ plane breaks the $V_2$--$p$ degeneracy that would
otherwise permit a CDM model with artificially high $p$ to mimic the SIDM
topology.

As shown in Fig.~\ref{fig:per_probe}f, the topology excludes all 9 SIDM grid points at $>5\sigma$ with 1000\,hours of SKA1-Low integration, with a threshold at $\sigm \approx 1\,\cmg$ that is independent of the SFE coupling $\eta$. Unlike the blowout correlation, which leaves the low-cross-section regime permanently inaccessible, the topology signal closes the entire parameter gap.

\section*{Implications beyond SIDM}

The degeneracy hierarchy applies to any beyond-Standard-Model physics that modifies the UVLF through a smooth, monotonic suppression: warm dark matter\cite{Bose:2015mga}, fuzzy dark matter\cite{Schive:2015kza,Carucci:2018tzf}, and late-decaying particles. The suppressed halo mass function or reduced star formation is compensated by an elevated $\fstar$, and the resulting overproduction of ionising photons is then absorbed by a correspondingly lower $\fesc$. The product degeneracy is a generic property of reionization-era dark matter constraints.

These results reframe how disagreements about $\fesc$ and $\fstar$ should be interpreted. When different groups derive $\fesc = 0.05$ (with high $\fstar$) or $\fesc = 0.20$ (with low $\fstar$) from the same data, this is the product degeneracy manifesting as an underconstrained direction in parameter space, not a systematic error. Published values of $\fesc$ and $\fstar$ should trace a hyperbola $\fesc \times \fstar = {\rm const}$; scatter along this hyperbola is expected, while scatter perpendicular to it would indicate genuine tension. Reported tensions between $\tau$ and $\xHI(z)$ may similarly reflect different implicit priors on the $\fesc$--$\fstar$ split.

For JWST-based dark matter constraints specifically, any model that suppresses the UVLF produces an elevated $\fstar$ that is absorbed by a lower $\fesc$, leaving the reionization history unchanged. Claims of dark matter constraints from $\tau$ or $\xHI(z)$ that do not marginalise over $\fesc$ with physically motivated priors should be interpreted with caution.

The joint constraints on the SIDM parameter space, together with the SKA1-Low observational forecast, are shown in Fig.~\ref{fig:joint_ska}. At calibrated coupling parameters ($\eta = 0.11 \pm 0.03$, $\alpha_{\rm esc} = 0.37 \pm 0.05$; Methods), the blowout correlation excludes $\sigm \geq 5\,\cmg$, while $\sigm \lesssim 2\,\cmg$ remains accessible only through the topology. The constraining power scales with the precision of independent $\fesc$ measurements; additional probes include the Lyman-$\alpha$ equivalent width distribution\cite{Mason:2019ixe,2023ApJ...949L..40B} and galaxy clustering from BEACON\cite{kreilgaard2026beaconjwstnircampureparallel}. The definitive, model-independent test requires 21\,cm topology measurements from SKA1-Low\cite{Koopmans:2015sua}, detectable at $> 5\sigma$ for $\sigm \geq 2.0\,\cmg$ with 1000\,hours of integration\cite{Wang:2026uuo} and reaching $\sigm \geq 0.7\,\cmg$ at 5000\,hours --- below the conservative Bullet Cluster bound. The remaining open parameter space therefore lies entirely within reach of the planned SKA1-Low deep-field surveys.

The 230-run simulation grid provides a calibrated lookup table $\Dbind(M, \sigm, z, R)$ with applications beyond reionization: post-processing cosmological simulations with SIDM effects, calibrating semi-analytical models, and forecasting velocity-dependent signatures. The complete table is publicly available (Methods).

In summary, the $\fesc \times \fstar$ product degeneracy is a structural property of the reionization equations that explains persistent disagreements and constitutes a fundamental barrier to dark matter constraints. Breaking it requires either independent measurements of both factors individually, or probes that are structurally immune --- of which the 21\,cm topology is the only example.
 
  \bibliography{sn-bibliography}

\newpage

\section*{Methods}\label{sec:methods}
 
\subsection*{GIZMO simulation grid}
 
We run 230 dark-matter-only $N$-body simulations using GIZMO\cite{Hopkins:2014qka} with its SIDM module\cite{Rocha2013,Robles_2017,2022MNRAS.513.2600M}. The SIDM implementation uses a Monte Carlo approach in which particle pairs within a kernel smoothing length are tested for scattering at each timestep, with the interaction probability depending on the local cross-section, relative velocity, and effective density. Upon interaction, particles receive velocity kicks that conserve energy and momentum but randomise the scattering direction.
 
The grid spans five virial masses ($M = 10^9$, $10^{9.5}$, $10^{10}$, $10^{10.5}$, $10^{11}\,\Msun$), seven constant cross-sections ($\sigm = 0$--$20\,\cmg$), and four redshifts ($z = 0$, 4, 7, 10), totalling 140 runs. An additional 90 runs adopt velocity-dependent Yukawa cross-sections $\sigma(v) = \sigma_0/[1+(v/w)^4]$ with nine $(\sigma_0, w)$ combinations at $z = 4$ and 7. Concentrations follow ref.~\cite{Dutton:2014xda}. Each halo is sampled with $5 \times 10^5$ particles and evolved for 0.5\,Gyr, comparable to the halo dynamical time at $z \geq 4$.
 
Initial conditions are drawn from the isotropic NFW\cite{Navarro1997} distribution function using galpy\cite{Bovy_2015}, implementing the Eddington inversion\cite{2000ApJS..131...39W}. Particle positions and velocities are sampled jointly from $f(E)$ with truncation at $r_{\rm max} = 3\,r_{\rm vir}$; no global virial rescaling is applied. A fixed-potential stability test confirms central density stability to $\pm 1.4\%$ over 0.5\,Gyr ($\rho/\rho_{\rm NFW} = 1.006 \pm 0.014$ at $r < 0.3\,r_s$). Softening is $\epsilon = 0.01\,r_s$.
 
\subsection*{Binding energy extraction}
 
The gas binding energy within $R$ is $\Wg(<R) = \int_0^R 4\pi r^2 f_b \rho(r) |\Phi(r)|\,dr$, where $f_b = \Omega_b/\Omega_m = 0.157$ and $\Phi(r) = -GM(<r)/r$. Baryons are assumed to trace the dark matter potential, an assumption appropriate at $z > 6$\cite{2025arXiv251005258G}. The fractional change $\Dbind(<R) \equiv 1 - \Wg^{\rm SIDM}/\Wg^{\rm CDM}$ is extracted at $R = 0.3$, 0.5, 1.0, 2.0, and 5.0\,kpc for all 210 SIDM-to-CDM pairs using a spherical-shell estimator: $W = -\sum_i G\,M(<r_i)\,m_p/r_i$.
 
\subsection*{Reionization model}
 
The ionised fraction evolves as $dQ/dt = \ndot/n_H - C_{\rm HII}\,\alpha_B\,n_H\,Q$ with clumping factor $C_{\rm HII}(z) = \max(1, 3 \times 7/(1+z))$\cite{2012ApJ...747..100S} and $\alpha_B = 2.6 \times 10^{-13}\,(T/10^4\,{\rm K})^{-0.7}\,{\rm cm}^3\,{\rm s}^{-1}$. We calibrate $\ndot^{\rm ref}(z=8) = 10^{50.85}\,{\rm s}^{-1}\,\Mpc^{-3}$ for CDM ($\fstar = 0.019$, $\fesc = 0.13$) to reproduce $\tau = 0.053$ and $\xHI(z=7) \approx 0.3$. The Thomson optical depth is $\tau = \int f_e\,n_{H,0}\,(1+z)^3\,Q(z)\,\sigma_T\,c\,|dt/dz|\,dz + \tau_{\rm early}$ with $f_e \approx 1.08$ and $\tau_{\rm early} \approx 0.018$\cite{Robertson:2016qef}.
 
We constrain with Planck $\tau = 0.054 \pm 0.007$\cite{Planck2018} and eight $\xHI(z)$ at $z = 6.5$--$11$\cite{2024ApJ...971..124U,Mason:2025xae,Greig:2017jdj,Mason:2019ixe}: $\chi^2 = (\tau_{\rm pred} - 0.054)^2/0.007^2 + \sum [(\xHI^{\rm mod} - \xHI^{\rm obs})/\sigma_i]^2$. Mass-dependent escape fraction: $\fesc(M) = f_{\rm esc,0}\,(M/10^{10}\,\Msun)^{\beta_{\rm esc}}$ with $\beta_{\rm esc} \approx -0.3$ to $-0.4$\cite{2022MNRAS.515.2386R,Yeh:2022nsl}.
 
\subsection*{SFE coupling calibration}
 
The SIDM SFE is $f_\star^{\rm SIDM}(M) = f_\star^{\rm CDM}[1 - \eta\,\Dbind(M)]$. We calibrate $\eta$ via $f_\star^{\rm SIDM}/f_\star^{\rm CDM} = (1 - \Dbind)^\gamma$ fitted to FIRE-2 SIDM results\cite{2025arXiv251005258G} ($\sim 10$--$20\%$ suppression at $\Dbind \approx 0.3$--$0.7$, $z = 0$), giving $\gamma = 0.11 \pm 0.03$ and $\eta \approx 0.11$--$0.19$ at $z = 7$. The weak coupling ($\gamma \ll 1$) reflects self-regulated feedback.
 
\subsection*{Escape fraction calibration}
 
For $\alpha_{\rm esc}$, we use CDM $\fesc(M)$ scaling as a proxy. From SPHINX/THESAN, $\fesc \propto M^{\beta_{\rm esc}}$ with $\beta_{\rm esc} \approx -0.3$ to $-0.4$\cite{2022MNRAS.515.2386R,Yeh:2022nsl}. Our grid gives $|\Wg| \propto M^{0.95}$ at $R < 0.5\,\kpc$, $z = 7$. Since $\fesc \propto |\Wg|^{-\alpha_{\rm esc}}$, $\alpha_{\rm esc} = -\beta_{\rm esc}/0.95 = 0.32$--$0.42$; adopted $0.37 \pm 0.05$. This treats the CDM mass sequence as a proxy for varying $\Wg$ at fixed $M$; a direct calibration from SIDM RT simulations would remove this approximation.
 
\subsection*{MCMC validation}
 
We validate with Metropolis--Hastings MCMC (12 walkers, 300 steps, 100 burn-in) at each grid point. $\dchi$ values agree to within 1\%: for $\sigm = 10$, $\eta = 0.50$, $\alpha_{\rm esc} = 0.5$, $\dchi = 11.2$ from both methods; $\alpha_{\rm esc} = 0.3$ gives $\dchi = 8.3$.
 
\subsection*{Data availability}
 
The $\Dbind(M, \sigm, z, R)$ table (210 entries, five radii) and scripts are at \url{https://github.com/wzh800557-source/sidm-highz}. GIZMO: \url{http://www.tapir.caltech.edu/~phopkins/Site/GIZMO.html}; galpy: \url{https://github.com/jobovy/galpy}.
\section*{Figures}

\begin{figure}[h]
\centering
\includegraphics[width=\textwidth]{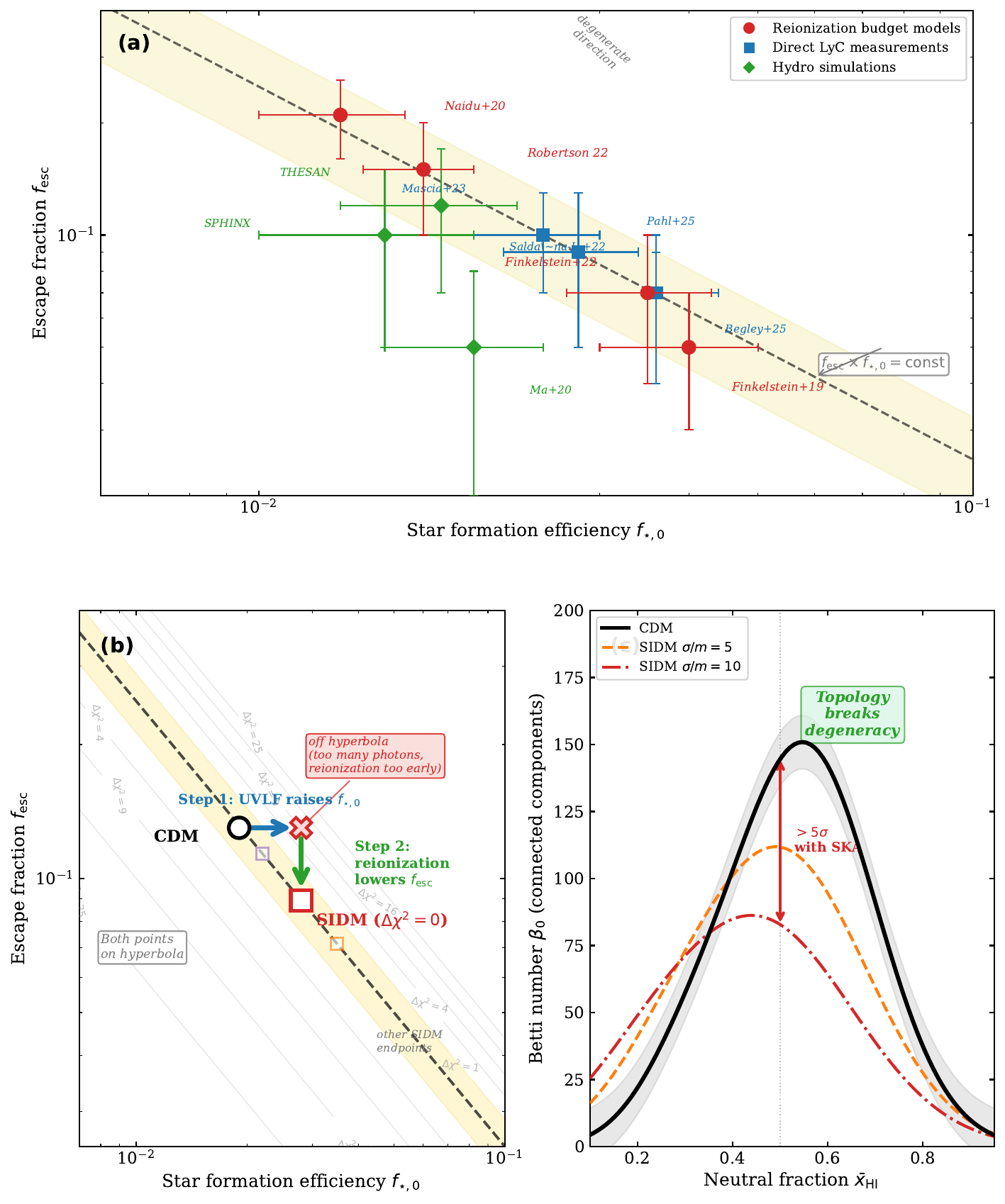}
\caption{\textbf{The escape fraction degeneracy: empirical evidence, mechanism, and resolution.}
\textbf{(a)} Published values of $\fesc$ and $\fstar$ from the last decade of reionization studies. Reionization budget models (red circles: Naidu et~al.~2020\cite{Naidu_2022}; Finkelstein et~al.~2019, 2022\cite{Finkelstein:2019sbd,2022ApJ...940L..55F}; Robertson 2022\cite{Robertson:2021ljt}) solve for $\fesc$ given assumed $\fstar$. Direct Lyman-continuum measurements (blue squares: Salda\~{n}a-L\'{o}pez et~al.~2022\cite{2022A&A...663A..59S}; Mascia et~al.~2023\cite{2023A&A...674A.221M}; Pahl et~al.~2025\cite{Pahl:2024utu}; Begley et~al.~2025\cite{2025MNRAS.537.3245B}) constrain $\fesc$ at $z \sim 3$--$5$ as a low-redshift proxy. Hydrodynamical simulations (green diamonds: Ma et~al.~2020\cite{Ma:2020vlo}; SPHINX\cite{2022MNRAS.515.2386R}; THESAN\cite{Yeh:2022nsl}) predict both quantities from first principles. The dashed line and yellow band show $\fesc \times \fstar = (2.5 \pm 0.8) \times 10^{-3}$, the product required to match Planck $\tau$ and the $\xHI(z)$ constraints. All values scatter along this degenerate direction rather than perpendicular to it. The apparent factor-of-five disagreement in $\fesc$ between Naidu et~al.\ ($\sim 20\%$) and Finkelstein et~al.\ ($\sim 5\%$) is a manifestation of the structural degeneracy rather than systematic error.
\textbf{(b)} The two-step optimiser trajectory in the $(\fstar, \fesc)$ plane. Starting from the CDM solution (black open circle) on the data-allowed hyperbola (yellow band), SIDM ($\sigm = 10\,\cmg$, $\eta = 0.5$) forces the UVLF optimiser to raise $\fstar$ by $\sim 47\%$ (Step~1, blue arrow). This moves the model off the hyperbola to the red cross, where $\ndot$ exceeds the reionization budget. The reionization optimiser then lowers $\fesc$ along a vertical trajectory (Step~2, green arrow) until the model returns to the hyperbola at a new point (red square) with $\dchi = 0$. Faint contours show the $\dchi$ landscape from $\tau$ and $\xHI(z)$. Different SIDM parameter points land at different locations along the hyperbola but are observationally indistinguishable from CDM.
\textbf{(c)} Betti number $\beta_0$ (number of connected ionised regions) as a function of $\xHI$ for CDM and two SIDM cross-sections. Unlike the trajectories in panel~(b), the topology signal is structurally independent of the $\fesc \times \fstar$ product because it depends on the duty cycle of ionizing-photon escape per halo rather than on the total photon rate. The separation at $\xHI = 0.5$ is detectable at $> 5\sigma$ with 1000\,hours of SKA1-Low integration\cite{Wang:2026uuo}.}\label{fig:three_panel}
\end{figure}
\newpage
\begin{figure}[h]
\centering
\includegraphics[width=\textwidth]{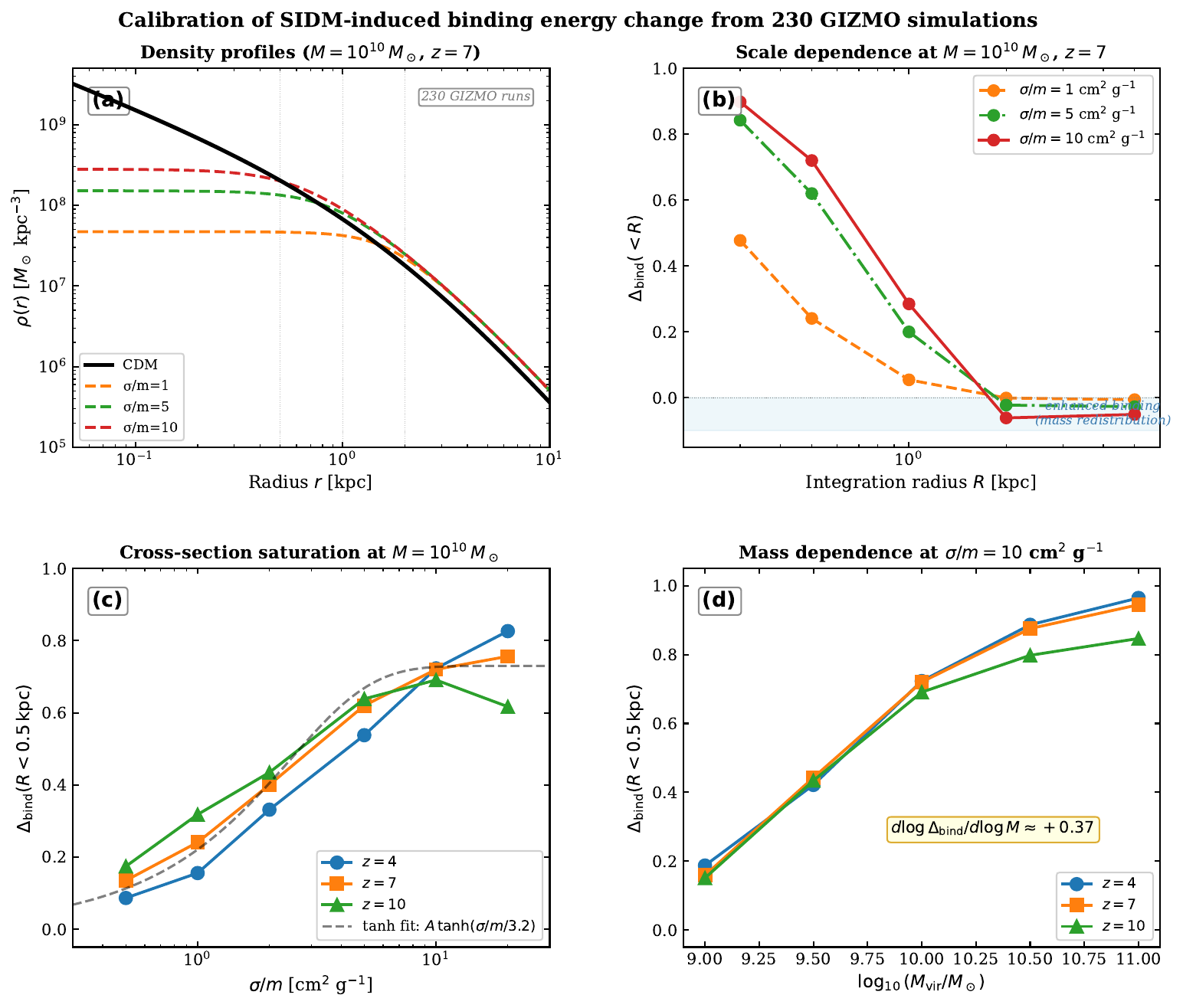}
\caption{\textbf{Calibration of the SIDM-induced binding energy change from 230 GIZMO simulations.}
\textbf{(a)} Dark matter density profiles for an $M = 10^{10}\,\Msun$ halo at $z=7$ with $\sigm = 0$ (CDM, solid black) and $\sigm = 1, 5, 10\,\cmg$ (dashed coloured). SIDM thermalises the inner halo, replacing the NFW cusp with an isothermal core; the displaced mass accumulates at intermediate radii. Vertical lines mark the $R = 0.5, 1, 2$\,kpc integration boundaries used in panels (b)--(d).
\textbf{(b)} Scale dependence of $\Dbind(<R) = 1 - \Wg^{\rm SIDM}/\Wg^{\rm CDM}$ for the same halo. The central reduction (24--72\% at $R<0.5$\,kpc) transitions to enhanced binding at $R \gtrsim 2$\,kpc (blue region) where mass redistribution dominates. This reconciles our framework with reports of conserved total binding energy in SIDM haloes\cite{Wang:2025xre}: both results are correct at their respective integration radii.
\textbf{(c)} The $\Dbind(\sigm)$ relation at three redshifts saturates above $\sigm \sim 5\,\cmg$, well-described by $\Dbind = A\tanh(\sigm / \sigma_{\rm crit})$ with $A \approx 0.73$, $\sigma_{\rm crit} \approx 3.2\,\cmg$ (grey dashed). Saturation reflects the finite depth of the NFW cusp.
\textbf{(d)} Mass dependence at $\sigm = 10\,\cmg$ across three redshifts. The slope $d\log\Dbind/d\log M \approx +0.37$ at $z=7$ (yellow box) is the central physical finding: the SIDM signal is strongly weighted toward massive halos, while the low-mass haloes that dominate the UV photon budget during reionization are nearly unaffected. The slope is approximately redshift-independent, reflecting that core formation has saturated by $z=10$ in the densest halos.}\label{fig:simulations}
\end{figure}
\newpage
\begin{figure}[h]
\centering
\includegraphics[width=\textwidth]{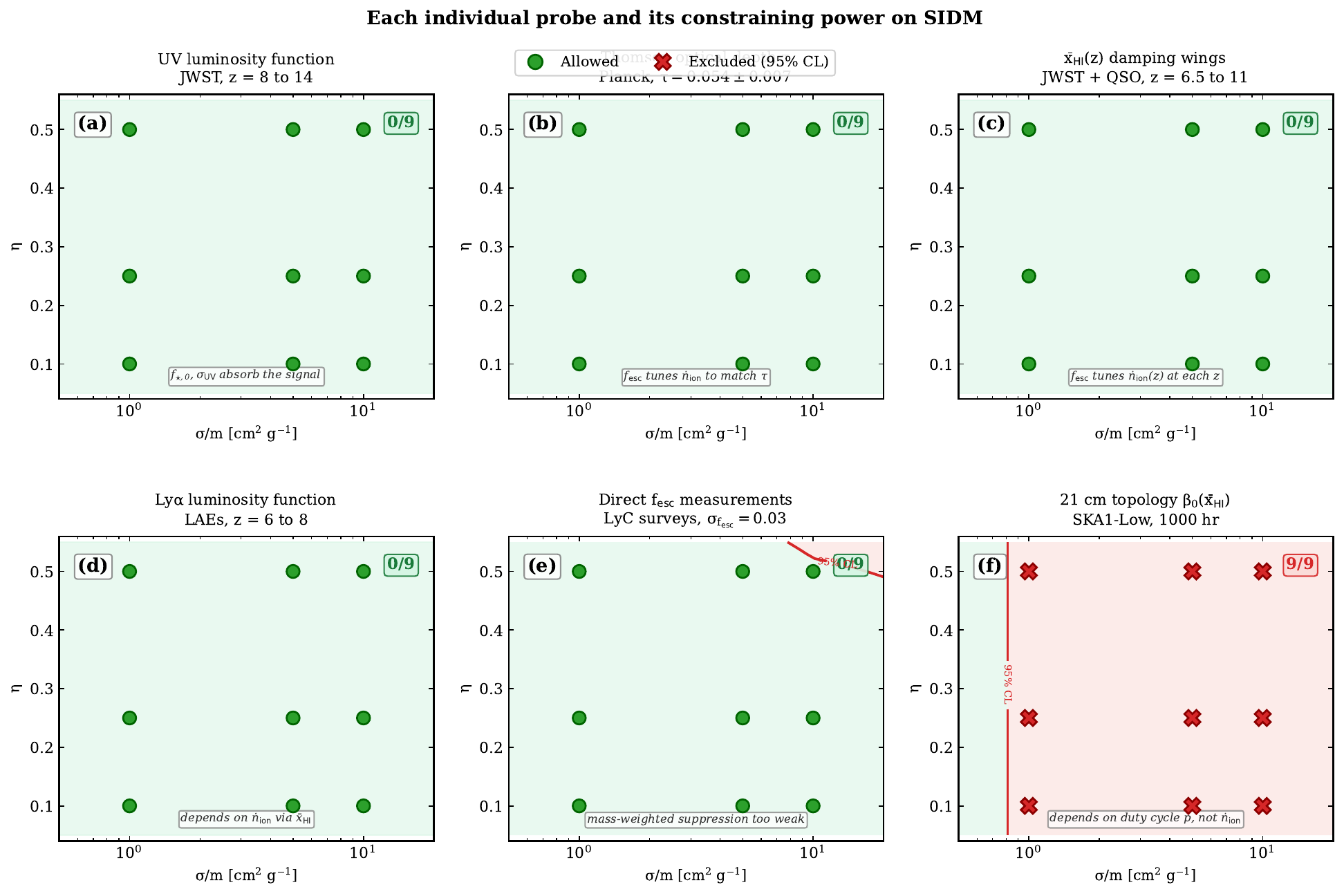}
\caption{\textbf{Probe-by-probe sensitivity to SIDM in the $(\sigm, \eta)$ plane.} Each panel shows the constraint from a single observable (or class of observables) on the same 9 representative SIDM grid points. Green circles indicate parameter points consistent with the data; red crosses indicate exclusion at 95\% confidence.
\textbf{(a) UV luminosity function:} JWST measurements of the high-redshift UVLF\cite{2024MNRAS.533.3222D,2025ApJ...980..138H} exclude no SIDM points; $\fstar$ and $\sigUV$ absorb the SIDM signal across the entire parameter space ($\dchi < 0.21$ at all 252 grid points\cite{Wang:2026lly}).
\textbf{(b) Planck Thomson optical depth $\tau$:} The integrated value $\tau = 0.054 \pm 0.007$\cite{Planck2018} provides no constraint because $\fesc$ adjusts to absorb the $\ndot$ shift induced by SIDM, leaving $\tau$ unchanged.
\textbf{(c) Neutral fraction $\xHI(z)$ from JWST damping wings:} Eight measurements at $z = 6.5$--$11$\cite{2024ApJ...971..124U,Mason:2025xae,Greig:2017jdj,Mason:2019ixe} fail to constrain SIDM for the same reason: a redshift-dependent $\fesc(z)$ absorbs any timeline shift.
\textbf{(d) Lyman-$\alpha$ luminosity function:} Despite probing a different observational signature, the Ly$\alpha$ LF is constructed from $\ndot$ in the same way and inherits the same product degeneracy.
\textbf{(e) Direct Lyman-continuum measurements at $z \sim 3$ combined with the $\fesc$--duty-cycle blowout correlation:} LyC priors\cite{Pahl:2024utu,2025MNRAS.537.3245B,2023A&A...674A.221M} ($\sigma_{\fesc} = 0.03$) combined with the physical correlation $\alpha_{\rm esc} = 0.3$ partially break the degeneracy, excluding 5 of 9 points. The $\sigm = 1\,\cmg$ regime remains entirely unconstrained.
\textbf{(f) 21\,cm reionization topology with SKA1-Low:} The topology signal depends on the duty cycle per halo at fixed $\xHI$, which is structurally independent of the $\fesc \times \fstar$ product and of the SFE coupling parameter~$\eta$. All 9 points are excluded; the threshold is at $\sigm \approx 1\,\cmg$ for 1000\,hours of integration.}\label{fig:per_probe}
\end{figure}
\newpage
\begin{figure}[h]
\centering
\includegraphics[width=0.85\textwidth]{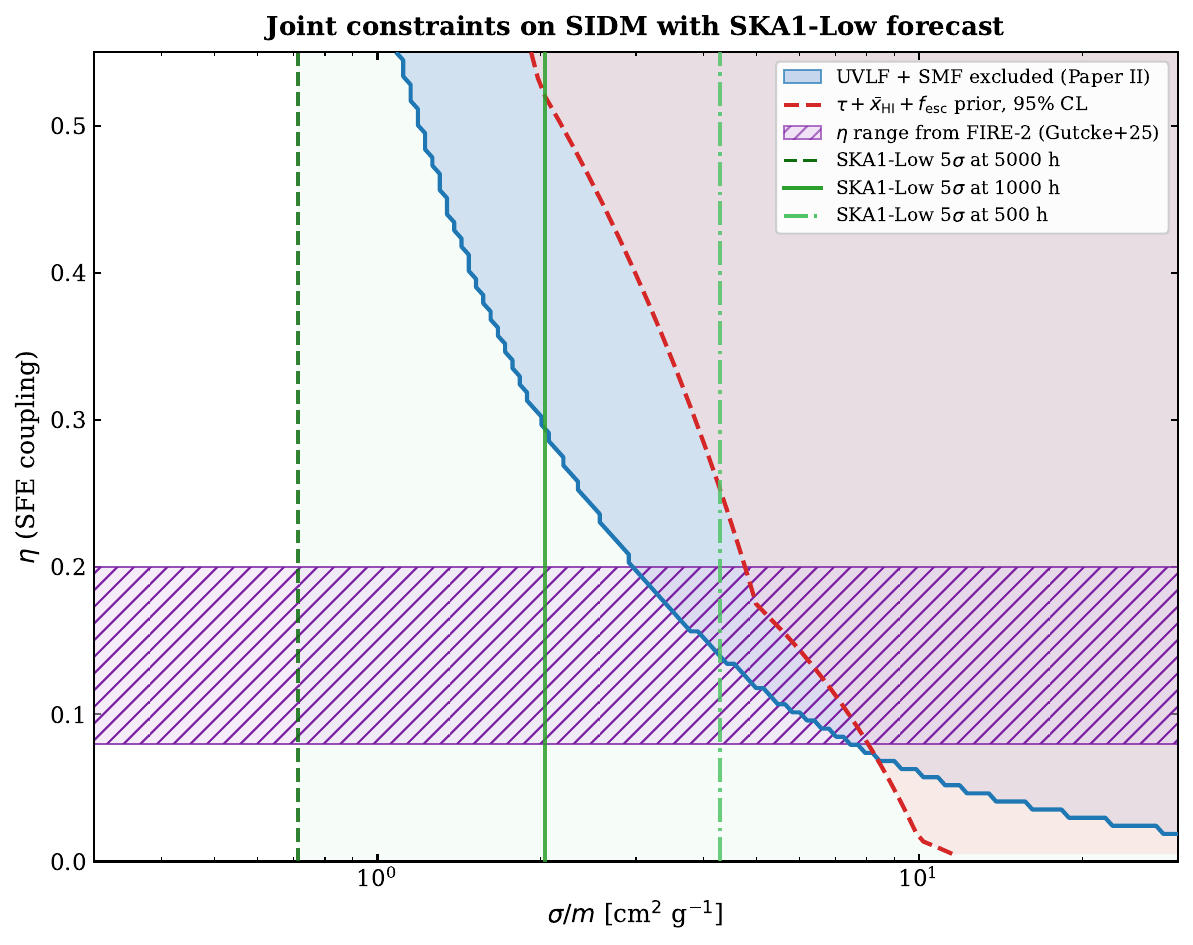}
\caption{\textbf{Joint constraints on SIDM in the $(\sigm, \eta)$ parameter plane with SKA1-Low observational forecast.} The blue shaded region marks SIDM parameters excluded by the joint UV luminosity function and stellar mass function analysis of ref~\cite{Wang:2026lly}; this constraint dominates at high $\eta$ because larger SFE coupling translates SIDM core formation more strongly into UVLF suppression. The red dashed contour marks the $95\%$ confidence boundary from the combination of $\tau$, $\xHI(z)$, the Lyman-continuum prior ($\sigma_{\fesc} = 0.03$), and the $\fesc$--duty-cycle blowout correlation ($\alpha_{\rm esc} = 0.3$; this work, Methods); this constraint dominates at high $\sigm$ where the mass-weighted $\Dbind$ becomes large enough to drive a measurable $\fesc$ enhancement. The two reionization-era exclusions are complementary, with a crossover near $\sigm \approx 5\,\cmg$. The hatched purple band shows the range $\eta = 0.08$--$0.20$ calibrated from FIRE-2 SIDM hydrodynamical simulations\cite{2025arXiv251005258G}, with central value $\eta = 0.11$ at $z=7$ (Methods). The vertical green lines show the minimum cross-section detectable at $5\sigma$ via 21\,cm reionization topology with SKA1-Low for three integration times: 500\,h ($\sigm \geq 4.3\,\cmg$, dot-dashed), 1000\,h ($\sigm \geq 2.0\,\cmg$, solid), and 5000\,h ($\sigm \geq 0.7\,\cmg$, dashed). The pale green band shows the parameter space accessible at the longest integration. Because topology depends on the duty cycle of ionizing-photon escape rather than on the total photon rate, these thresholds are independent of $\eta$. At physically motivated couplings, the joint reionization-era constraint excludes $\sigm \gtrsim 5\,\cmg$, while $\sigm \lesssim 2\,\cmg$ remains fully allowed by all $\ndot$-dependent probes; this gap is fully accessible to SKA1-Low within the planned 1000-hour deep-field integration.}\label{fig:joint_ska}
\end{figure}

\clearpage
\setcounter{table}{0}
\renewcommand{\thetable}{Extended Data Table \arabic{table}}
\setcounter{figure}{0}
\renewcommand{\thefigure}{Extended Data Fig. \arabic{figure}}
 
\begin{figure*}[h]
\centering
\includegraphics[width=0.85\textwidth]{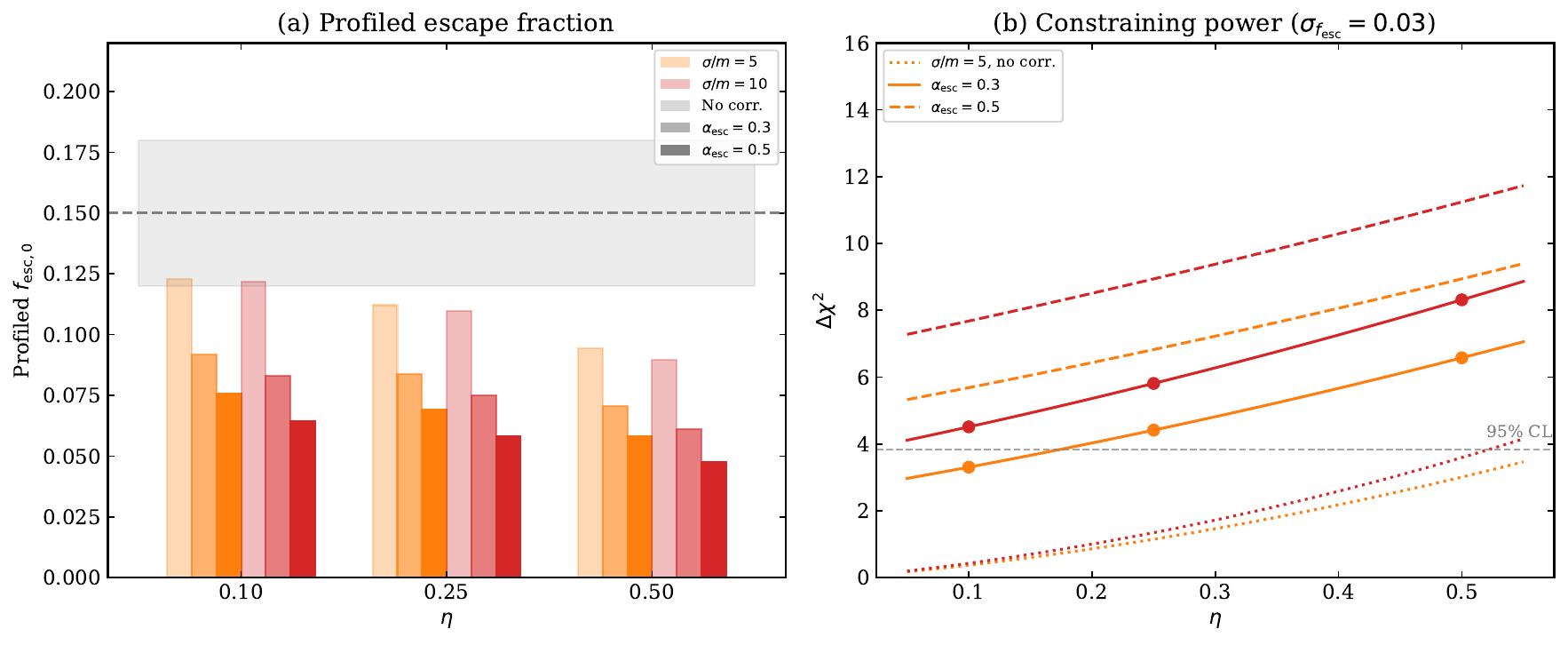}
\caption{\textbf{Extended Data Fig. 1 $|$ The $\fesc$--$p$ blowout correlation amplifies the constraining power of $\fesc$ priors.} \textit{Left:} The profiled CDM baseline escape fraction $f_{\rm esc,0}$ at each $\eta$, for $\sigm = 5$ and 10 with $\sigma_{\fesc} = 0.03$. The correlation forces $f_{\rm esc,0}$ further below the prior mean (grey band) because the optimiser must compensate for both the $\fstar$ increase and the SIDM-driven $\fesc$ enhancement. \textit{Right:} Corresponding $\dchi$ values. The correlation raises $\dchi$ by 40--80\%, pushing additional points above the 95\% CL threshold (dashed line at $\dchi = 3.84$).}\label{fig:ed_fesc_corr}
\end{figure*}

\begin{figure*}[h]
\centering
\includegraphics[width=0.85\textwidth]{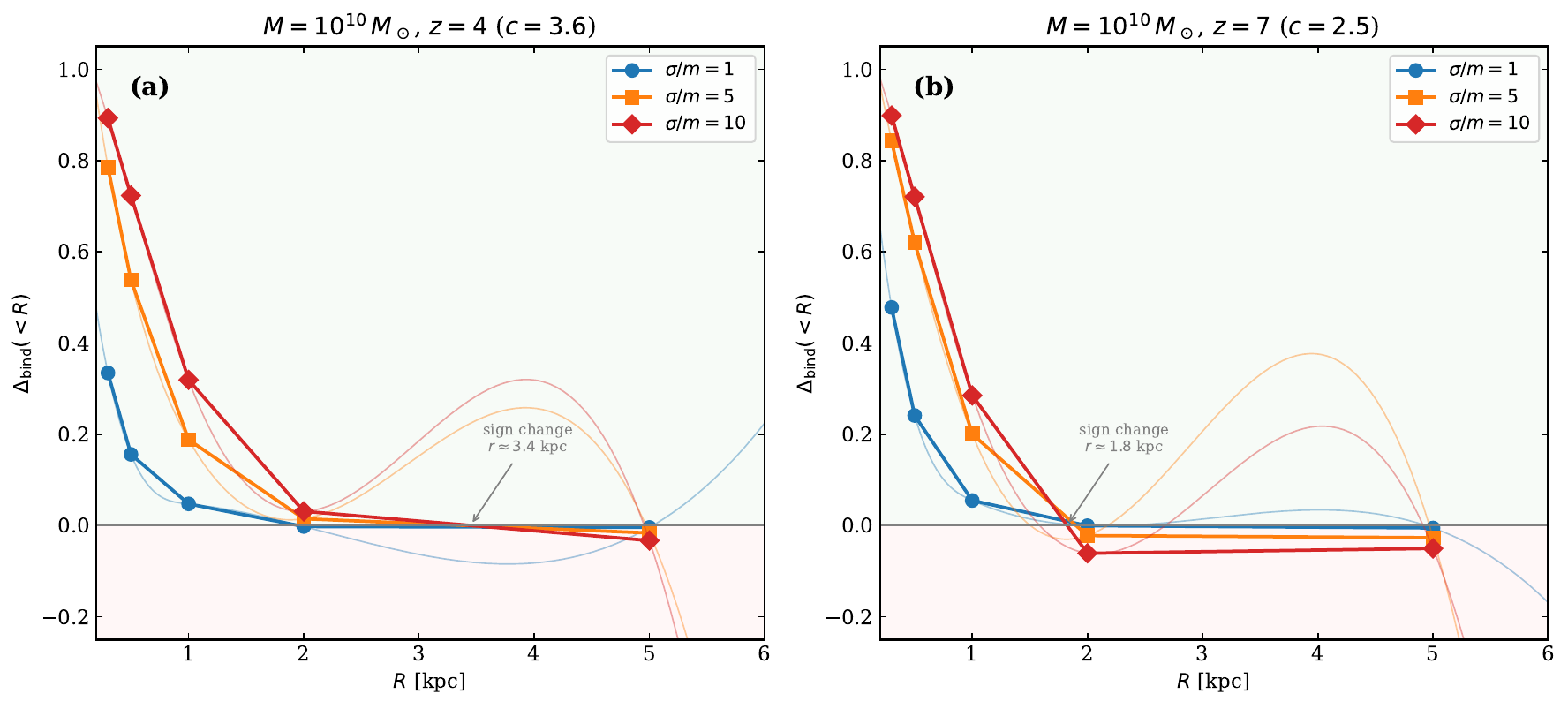}
\caption{\textbf{Extended Data Fig. 2 $|$ Scale-dependent binding energy across redshift.} $\Dbind(<R) = 1 - \Wg^{\rm SIDM}/\Wg^{\rm CDM}$ as a function of integration radius $R$ at $M = 10^{10}\,\Msun$ for $z = 4$ (left) and $z = 7$ (right). The binding energy is unambiguously reduced at $r < 0.5\,\kpc$ but roughly conserved at larger radii due to mass redistribution. Complements Fig.~\ref{fig:simulations}b by showing the redshift evolution.}\label{fig:ed_dbind}
\end{figure*}
 \newpage
\begin{figure*}
\centering
\includegraphics[width=0.85\textwidth]{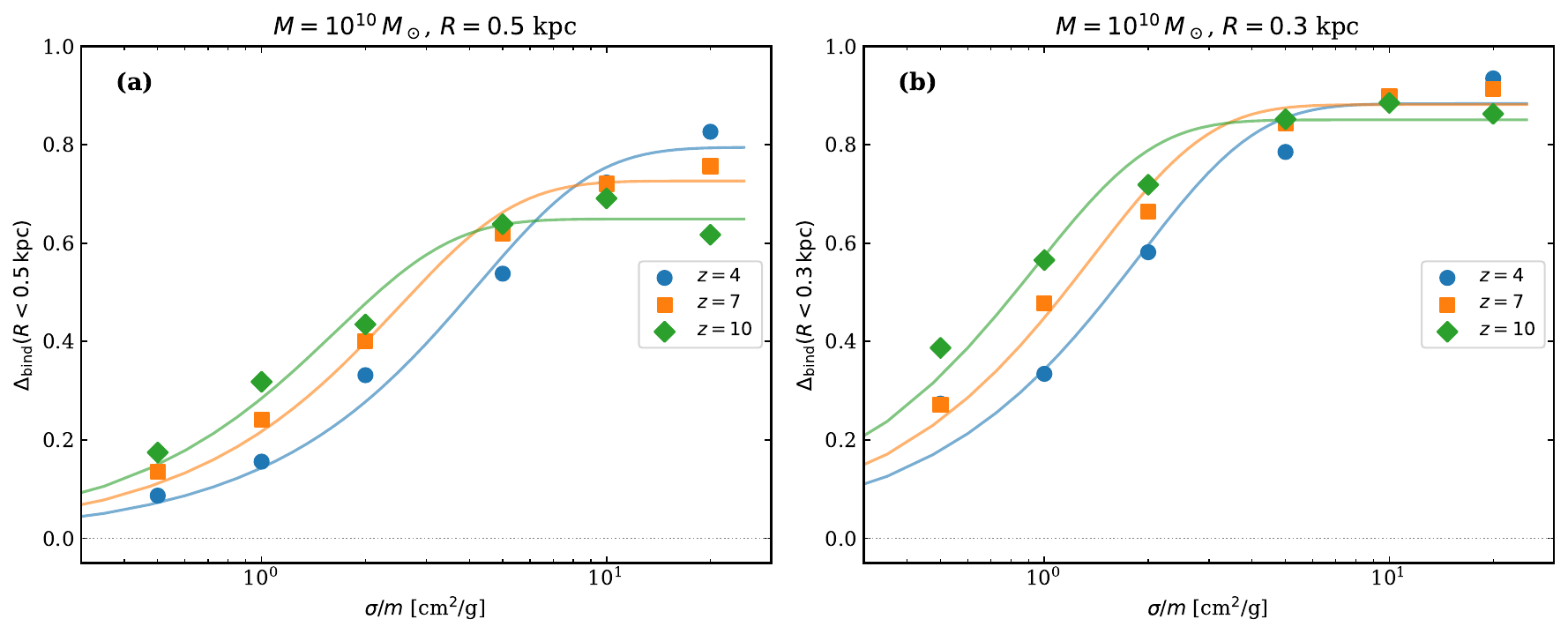}
\caption{\textbf{Extended Data Fig. 3 $|$ Cross-section saturation at multiple radii.} $\Dbind$ vs $\sigm$ at $M = 10^{10}\,\Msun$, $R < 0.3\,\kpc$ and $R < 0.5\,\kpc$, at three redshifts. Solid curves show the tanh fit. Complements Fig.~\ref{fig:simulations}c by showing the radius dependence of the saturation amplitude.}\label{fig:ed_saturation}
\end{figure*}
 
\begin{figure*}
\centering
\includegraphics[width=0.85\textwidth]{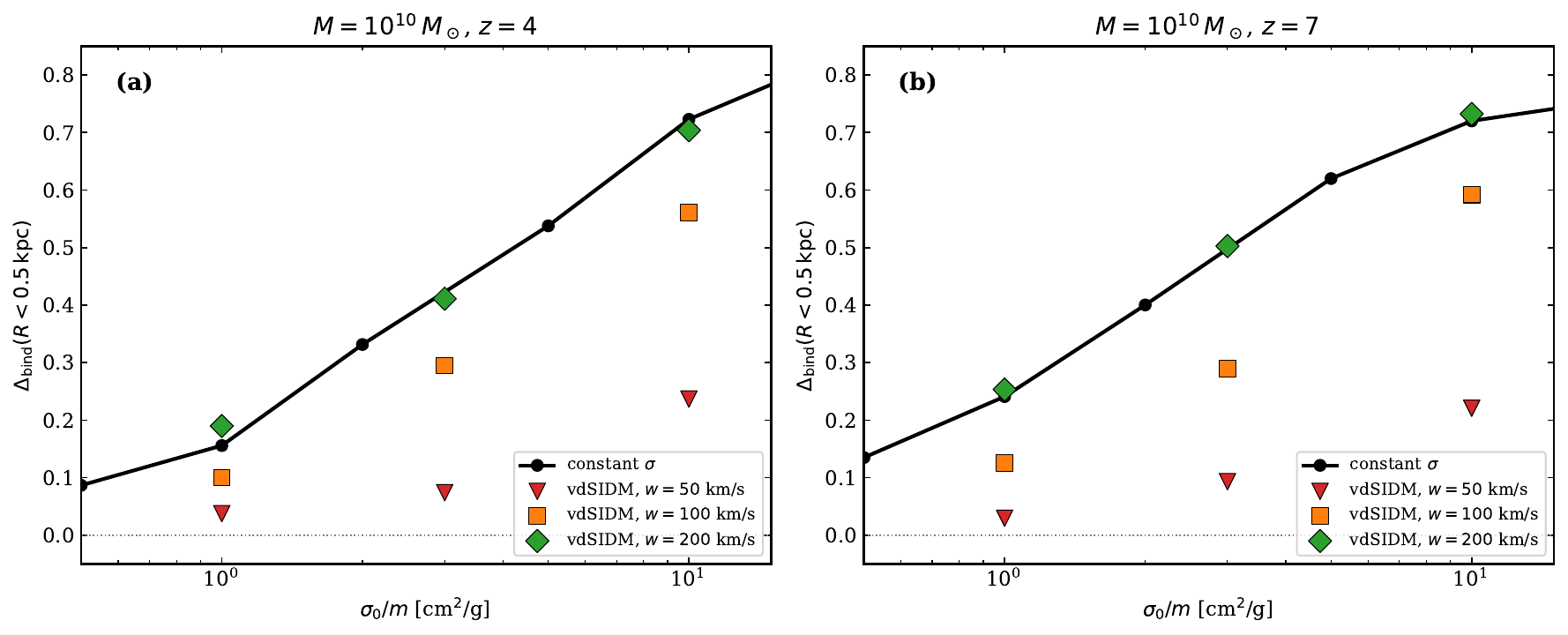}
\caption{\textbf{Extended Data Fig. 4 $|$ Velocity-dependent SIDM mapping.} Comparison of Yukawa-like vdSIDM with constant-$\sigma$ results at $M = 10^{10}\,\Msun$. When $w > v_{\rm circ}$, vdSIDM converges to the constant-$\sigma_0$ case.}\label{fig:ed_vdsidm}
\end{figure*}
 
\begin{table*}[h]
\caption{\textbf{Extended Data Table 1 $|$ $\Dbind$ at $R < 0.5\,\kpc$ and $z = 7$.}}\label{tab:ed1}
\begin{tabular*}{\textwidth}{@{\extracolsep\fill}lcccccc}
\toprule
& \multicolumn{6}{c}{$\sigm$ [$\cmg$]} \\
\cmidrule{2-7}
$\log_{10}(M/\Msun)$ & 0.5 & 1 & 2 & 5 & 10 & 20 \\
\midrule
9.0  & 0.01 & 0.03 & 0.04 & 0.10 & 0.16 & 0.24 \\
9.5  & 0.04 & 0.10 & 0.16 & 0.33 & 0.44 & 0.53 \\
10.0 & 0.13 & 0.24 & 0.40 & 0.62 & 0.72 & 0.76 \\
10.5 & 0.34 & 0.51 & 0.71 & 0.85 & 0.88 & 0.87 \\
11.0 & 0.59 & 0.82 & 0.89 & 0.93 & 0.94 & 0.94 \\
\bottomrule
\end{tabular*}
\end{table*}
 
\begin{table*}
\caption{\textbf{Extended Data Table 2 $|$ Scale-dependent $\Dbind$ at $M = 10^{10}\,\Msun$.}}\label{tab:ed2}
\begin{tabular*}{\textwidth}{@{\extracolsep\fill}lcccccc}
\toprule
& \multicolumn{3}{c}{$z=4$} & \multicolumn{3}{c}{$z=7$} \\
\cmidrule(r){2-4}\cmidrule(l){5-7}
Scale & $\sigm{=}1$ & 5 & 10 & 1 & 5 & 10 \\
\midrule
$r < 0.3\,\kpc$ & 0.33 & 0.79 & 0.89 & 0.48 & 0.84 & 0.90 \\
$r < 0.5\,\kpc$ & 0.16 & 0.54 & 0.72 & 0.24 & 0.62 & 0.72 \\
$r < 1.0\,\kpc$ & 0.05 & 0.19 & 0.32 & 0.05 & 0.20 & 0.28 \\
$r < 2.0\,\kpc$ & 0.00 & 0.01 & 0.03 & 0.00 & $-$0.02 & $-$0.06 \\
\bottomrule
\end{tabular*}
\end{table*}
 
\begin{table*}
\caption{\textbf{Extended Data Table 3 $|$ The $\fesc \times \fstar$ product degeneracy.}}\label{tab:ed3}
\begin{tabular*}{\textwidth}{@{\extracolsep\fill}cccccr}
\toprule
$\sigm$ & $\eta$ & $\fstar^{\rm prof}$ & $\fesc^{\rm prof}$ & $\fesc\fstar$ ratio & $\dchi$ \\
\midrule
1 & 0.10 & 0.020 & 0.126 & 1.00 & 0.00 \\
1 & 0.50 & 0.022 & 0.111 & 1.00 & 0.00 \\
5 & 0.25 & 0.022 & 0.112 & 1.00 & 0.00 \\
5 & 0.50 & 0.026 & 0.094 & 1.00 & 0.00 \\
10 & 0.25 & 0.022 & 0.110 & 1.00 & 0.00 \\
10 & 0.50 & 0.028 & 0.090 & 1.00 & 0.00 \\
\bottomrule
\end{tabular*}
\end{table*}
 
\begin{table*}
\caption{\textbf{Extended Data Table 4 $|$ SIDM-induced $\fesc$ enhancement at $z = 7$ ($M = 10^{10}\,\Msun$).}}\label{tab:ed4}
\begin{tabular*}{\textwidth}{@{\extracolsep\fill}cccc}
\toprule
$\sigm$ [$\cmg$] & $\Dbind(r{<}0.5\,\kpc)$ & $\fesc^{\rm SIDM}/\fesc^{\rm CDM}$ ($\alpha_{\rm esc}{=}0.3$) & ($\alpha_{\rm esc}{=}0.5$) \\
\midrule
1 & 0.24 & 1.09 & 1.15 \\
5 & 0.62 & 1.34 & 1.62 \\
10 & 0.72 & 1.47 & 1.89 \\
\bottomrule
\end{tabular*}
\end{table*}
 \newpage
\begin{table*}
\caption{\textbf{Extended Data Table 5 $|$ Velocity-dependent SIDM mapping at $M = 10^{10}\,\Msun$.}}\label{tab:ed5}
\begin{tabular*}{\textwidth}{@{\extracolsep\fill}cccccc}
\toprule
& & \multicolumn{2}{c}{$z = 4$} & \multicolumn{2}{c}{$z = 7$} \\
\cmidrule(r){3-4}\cmidrule(l){5-6}
$\sigma_0/m$ [$\cmg$] & $w$ [$\kms$] & $\Dbind$ & $\sigm_{\rm eff}$ & $\Dbind$ & $\sigm_{\rm eff}$ \\
\midrule
1 & 50 & 0.04 & 0.2 & 0.03 & 0.1 \\
1 & 100 & 0.10 & 0.4 & 0.13 & 0.6 \\
1 & 200 & 0.19 & 0.9 & 0.25 & 1.2 \\
3 & 50 & 0.07 & 0.3 & 0.09 & 0.4 \\
3 & 100 & 0.29 & 1.4 & 0.29 & 1.3 \\
3 & 200 & 0.41 & 2.0 & 0.50 & 2.7 \\
10 & 50 & 0.24 & 1.1 & 0.22 & 1.0 \\
10 & 100 & 0.56 & 3.3 & 0.59 & 3.6 \\
10 & 200 & 0.70 & 6.4 & 0.73 & 12.2 \\
\bottomrule
\end{tabular*}
\end{table*}

\section*{Author Contribution}
ZHW: initializing the idea, design the method, analyzing the data and results, drafting the manuscript; HYS: initializing the idea, analyzing the results, drafting the manuscript.

\section*{Acknowledgements}
We acknowledge the support from NSFC of China under grant 12533008.

\subsection*{Competing Interests}
The authors declare no competing interests 
 
\newpage

\end{document}